# Engineering thermal transport across layered graphene-MoS$_2$ superlattices

Aditya Sood[1,2], Charles Sievers[3,#], Yong Cheol Shin[1,#], Victoria Chen[1], Shunda Chen[3], Kirby K. H. Smithe[1], Sukti Chatterjee[4], Davide Donadio[3,*], Kenneth E. Goodson[5,6], Eric Pop[1,6,*]

[1]Department of Electrical Engineering, Stanford University, Stanford CA 94305
[2]Present address: Stanford Institute for Materials and Energy Sciences, SLAC National Accelerator Laboratory, Menlo Park CA 94025
[3]Department of Chemistry, University of California, Davis CA 95616
[4]Applied Materials Inc., Santa Clara CA 95054
[5]Department of Mechanical Engineering, Stanford University, Stanford CA 94305
[6]Department of Materials Science and Engineering, Stanford University, Stanford CA 94305
[#]equal contribution
[*]Corresponding authors: ddonadio@ucdavis.edu, epop@stanford.edu

**Layering two-dimensional van der Waals materials provides unprecedented control over atomic placement, which could enable tailoring of vibrational spectra and heat flow at the sub-nanometer scale. Here, using spatially-resolved ultrafast thermoreflectance and spectroscopy, we uncover the design rules governing cross-plane heat transport in superlattices assembled from monolayers of graphene (*G*) and MoS$_2$ (*M*). Using a combinatorial experimental approach, we probe nine different stacking sequences: *G, GG, MG, GGG, GMG, GGMG, GMGG, GMMG, GMGMG* and identify the effects of vibrational mismatch, interlayer adhesion, and junction asymmetry on thermal transport. Pure *G* sequences display signatures of quasi-ballistic transport, whereas adding even a single *M* layer strongly disrupts heat conduction. The experimental data are described well by molecular dynamics simulations which include thermal expansion, accounting for the effect of finite temperature on the interlayer spacing. The simulations show that a change of only 1.5% in the layer separation can lead to a nearly 100% increase of the thermal resistance. Using these design rules, we experimentally demonstrate a 5-layer *GMGMG* superlattice with an ultralow effective cross-plane thermal conductivity comparable to air, paving the way for a new class of thermal metamaterials with extreme properties.**

2D materials | van der Waals | heterostructure | phonon | thermal boundary resistance | time-domain thermoreflectance

Controlling thermal transport at the nanoscale is a major engineering challenge with applications in nanoelectronics, photonics, and energy conversion. Traditional approaches of controlling heat flow have relied on *top-down* methods, where the thermal conductivity of a material is tuned either through the incorporation of defects[1,2] or *via* a reduction in dimensionality.[3] While these approaches have resulted in the discovery of new regimes of thermal transport, as well as novel applications, they offer limited flexibility in terms of the range of thermal properties that can be accessed. In contrast, two-dimensional (2D) materials offer the unique ability to engineer thermal transport in a *bottom-up* manner. Layer-by-layer (LBL) assembly enables the creation of synthetic heterostructures with artificially tailored optical and electronic properties.[4] Because the distribution of atomic masses and bond strengths can be varied on the length scale of individual atoms, it has also been suggested that phonon spectra and thermal transport can be engineered in extreme ways.[5,6] This tunability arises from variations in atomic composition on the scale of phonon wavelengths (few nanometers), as well as large coherence lengths of heat-carrying phonons traveling across van der Waals (vdW) interfaces.[7]

The ability to choose different 2D layers and stack them in a deterministic fashion opens a large range of possible metamaterial architectures with tailored thermal transport and thermoelectric conversion properties.[8] At the same time, understanding the factors limiting thermal transport at vdW heterointerfaces is crucial for several applications in photonics and nanoelectronics. Energy dissipation creates heat, and interfaces are often the primary bottleneck for cooling a device.[9,10] Using techniques like Raman spectroscopy, time-domain thermoreflectance (TDTR), and time-resolved x-ray diffraction, previous studies have characterized the thermal boundary resistance (TBR) of various vdW interfaces between 2D materials and substrates, as well as between different 2D layers.[11–18] It has generally been found that these interfaces are more resistive than those between isotropic, three-dimensional (3D) materials.[18]

Despite these efforts, much remains to be understood about the fundamental mechanisms governing heat transport in vdW superlattices, and a systematic understanding is presently lacking. To design a new class of vdW thermal metamaterials, as well as understand factors limiting heat dissipation in 2D heterostructure devices, the following key questions need to be addressed: What are the primary factors governing thermal transport at vdW junctions, in particular, what are the roles of vibrational mismatch and interlayer separation? How does heat flow across a *hetero*junction compare to that across a *homo*junction? Finally, can we create 3D solids with tailored thermal conductivity by stacking 2D materials with matched or mismatched vibrational modes, for a variety of applications?

To shed light on these questions, here we design an array of LBL-assembled vdW solids made of two dissimilar 2D materials: monolayer graphene (*G*) and monolayer MoS$_2$ (*M*). Through combinatorial stacking of these monolayers, we construct nine sequences, *G*, *GG*, *MG*, *GGG*, *GMG*, *GGMG*, *GMGG*,

*GMMG*, *GMGMG* (see **Fig. 1**), and measure their cross-plane thermal resistance, $R_{2D}$. This is done using a high-throughput experimental approach employing spatially-correlated TDTR microscopy[2,19] and optical/spectroscopic imaging. Combining experimental results with non-equilibrium molecular dynamics (NEMD), we investigate the effect of layer number (*G* vs. *GG* vs. *GGG* and *M* vs. *MM*) and vibrational mismatch (*GM* vs. *GG* and *MM*) on thermal transport. NEMD simulations with finite-temperature thermal expansion effects accurately predict the thermal resistances of the *G*, *GG*, *GGG*, *GMG* and *GMMG* sequences to within 20% of experimental values. The prediction for the thickest sequence (*GMGMG*) is ~30% lower than measured, but it becomes consistent if a larger-than-equilibrium interlayer separation arising from stacking disorder is considered. Taken together, our results establish design rules governing heat transport in vdW metamaterials. By exploiting these rules, we create a 9-atom thick 'artificial solid' consisting of a graphene/MoS$_2$ superlattice with an *effective* cross-plane thermal conductivity lower than air, making it one of the best-known thermal insulators among fully-dense materials.

**Results and Discussion**

MoS$_2$ and graphene monolayers were grown by chemical vapor deposition (CVD) (see Methods for details).[20,21] Heterostructures were assembled by transferring these layers using a polymer-based process onto SiO$_2$ (90 nm) on Si substrates (see Methods). These substrates were pre-patterned with ~50 nm thick Ti/Au alignment markers. Three samples were prepared in all, as follows. Samples D1 and D2 each have up to 4 layers, containing regions with *G*, *GG*, *MG*, *GMG* and *GMMG* stacking. Both samples were prepared from the same transfer process and were cleaved from a single chip within ~1 cm of each other. Sample D3 contains up to 5 layers, with regions of *GG*, *GGG*, *GMG*, *GGMG*, *GMGG*, and *GMGMG* stacking. Both D2 and D3 were annealed at 350°C for 3 hours under vacuum (pressure ≈7 μTorr) to investigate the effect of interlayer coupling on thermal transport, whereas D1 was not annealed.

**Figure 1A** shows a schematic of the various stacking sequences grouped by the number of layers *n*. These are: (*n* = 1) *G*; (*n* = 2) *GG*, *MG*; (*n* = 3) *GGG*, *GMG*; (*n* = 4) *GGMG*, *GMGG*, *GMMG*; and (*n* = 5) *GMGMG*. A cross sectional high-resolution transmission electron microscopy (HRTEM) image of the *GMMG* stack from the annealed sample D2 is shown in **Fig. 1B**. Four layers are distinguishable, with a total stack thickness of ≈1.7 nm between the top and bottom graphene. **Fig. 1C** shows a high-angle annular dark-field scanning transmission electron microscopy (HAADF-STEM) image, where atomic number contrast enables visualization of the two MoS$_2$ layers. From this image, we measure the *M-M* distance to be 0.6 – 0.7 nm, which is compatible with the 0.616 nm c-axis lattice parameter[22] in bulk 2H-MoS$_2$, indicating good interface quality in our LBL-assembled samples.

To measure the thermal resistance across the stacks, $R_{2D}$, we used TDTR microscopy (see Methods, **Fig. S1**-**S3**). The samples were capped with an 80 nm thick Al film using electron-beam (e-beam) evaporation to serve as a transducer (**Fig. S4**). The edges of the pre-patterned Ti/Au markers remained visible under an integrated dark-field microscope even after Al deposition, enabling coarse alignment of the sample under the TDTR laser.[7] Finer alignment was achieved by scanning the sample and constructing thermal resistance maps with a pixel size of 0.5 or 1 μm. The full-width at half-maximum (FWHM) spatial resolution of this technique is ≈2.2 μm (**Fig. S5**). By comparing the TDTR images with optical micrographs taken before Al capping, a one-to-one mapping was made between the stack configuration and measured $R_{2D}$. As mentioned before, a unique feature of our experiments is the ability to measure multiple stack configurations within the same sample. This is enabled by the finite size (about tens of μm) of the triangular MoS$_2$ monolayers and by tears (missing regions) in the graphene monolayers. Because we probe different regions within a few microns of one another, we can mitigate concerns about spatial variations in the quality of the top (between Al and the top 2D layer) and bottom interfaces (between the bottom 2D layer and SiO$_2$ substrate) which could otherwise affect the measured $R_{2D}$. For 8 of the 9 stacks, the top-most and bottom-most layers are *G*, which enables direct comparisons of the 'intrinsic' stack resistances.

**Fig. 2A** shows an optical image of a region of sample D1. Three stack sequences are distinguishable based on their optical contrast: *GG*, *GMG* and *GMMG*. MoS$_2$ photoluminescence (PL) emission measurements made on the same region and are shown in the inset of **Fig. 2B**; the maps plot the intensity integrated over the energy range 1.82 to 1.9 eV. The PL intensity of the *MM* region is brighter than monolayer *M* regions, suggesting weak interlayer electronic coupling. As has been shown previously, monolayer MoS$_2$ is a direct band gap semiconductor with bright PL emission, whereas bilayer MoS$_2$ (even with an arbitrary twist angle) is indirect band gap with significantly lower PL intensity.[23,24] In the as-prepared sample D1, the MoS$_2$ layers are weakly coupled and they behave as individual monolayers whose intensity is approximately doubled in the bilayer region.

**Fig. 2B** displays a TDTR microscopy image of the same region where we plot spatial distribution of $R_{2D}$. Details of the mapping technique are provided in the Methods section. A clear correspondence is seen between the optical, PL, and TDTR images. This visual correlation enables a direct extraction of $R_{2D}$ for the different stacking configurations without any image processing. The *GMMG* region is the most thermally resistive, followed by *GMG* and *GG*. The sample also has regions of *G* and *MG* stacking, which are not shown in this map. Note that the TDTR measurements were made after the sample was coated with an optically-opaque 80 nm thick layer of Al; therefore, these results demonstrate the sub-surface imaging capability enabled by TDTR microscopy. Single-spot TDTR time scans (over the full range of delays) for each stack configuration are provided in **Fig. 2C**. The decay rate of the ratio (= -$V_{in}$/$V_{out}$) signal encodes information about the interface thermal resistance: a faster (slower) rate indicates a lower (higher) $R_{2D}$. The interfaces show uniformity over large areas, as illustrated by the thermal resistance profile in **Fig. 2D**, which is plotted along the line marked in **Fig. 2B**.

In LBL-assembled vdW stacks, it has been suggested that the interface quality can be improved significantly and made 'intrinsic' by high-temperature annealing.[25,26] To probe its effect on interlayer thermal transport, we measure $R_{2D}$ on the annealed sample D2. **Fig. 3A** shows an optical image of a D2 region where five different regions are identified besides bare SiO$_2$: *G*, *GG*, *MG*, *GMG*, and *GMMG*. As discussed above, PL intensity provides a useful means to probe the strength of interlayer electronic coupling. We perform PL mapping of the same sample region (indicated by dash-dotted lines in **Fig. 3A**) before and after the anneal. As demonstrated in **Fig. 3B**, after annealing, there is a crossover in the MoS$_2$ PL intensity with *GMMG* showing quenched emission compared to *GMG*, consistent with stronger *M-M* electronic coupling. We confirm that this quenching is not due to enhanced *G-M* electronic coupling, as the PL intensity of *GMG* does not change significantly on annealing. This is also seen clearly in the PL point spectra in **Fig. S6**.

We employ Raman spectroscopy to probe the effect of annealing on interlayer vibrational and electronic coupling. **Fig. 3C** shows that, on annealing, the A$_1$' peak of MoS$_2$ in the *GMMG* region blueshifts, while the E' peak shows little change. This increased separation by ≈2 cm$^{-1}$ between the A$_1$' and E' phonon frequencies indicates stronger *M-M* vibrational coupling in the annealed sample.[26] **Fig. 3D** plots the Raman spectra of graphene in the *GG* region before and after annealing. We observe an increase in the FWHM of the 2D peak, by ≈6 cm$^{-1}$. This is consistent with improved *G-G* electronic coupling, and a modification of the electronic band structure in bilayer graphene.[27] The reduced intensity ratio of 2D and G peaks, from ~2.5 to ~1.8 further supports the notion of enhanced *G-G* interaction upon annealing.

The sample's thermal resistance map after annealing is shown in **Fig. 3E**, from which we extract $R_{GMMG}$, $R_{GMG}$, $R_{GG}$ and $R_G$. In **Fig. 3F** we plot the thermal resistances of the different stack sequences from samples D1 and D2 before and after annealing; these data are extracted by fitting single-spot TDTR time scans over the full delay range. To demonstrate the spatial uniformity of the interfaces, we plot histograms of $R_{2D}$ by defining regions of interest (ROI) in the TDTR maps, as shown in **Fig. 3G** (locations of ROI polygons are given in **Fig. S7** and **Fig. S8**). We find good agreement between the thermal resistance extracted from the full-delay-time single-spot scans and the fixed-delay-time 2D maps. Spatial variations are relatively small and do not affect our main conclusions. In the unannealed sample D1, we find that $R_{2D}$ scales approximately linearly with the number of layers (*n*), such that $R_{GMMG} > R_{GMG} > R_{GG} > R_G$. In comparison, in the annealed sample D2, $R_G$ is similar to D1, but $R_{GG}$ is reduced strongly by ~40%, such that $R_G \approx R_{GG}$ after the anneal. $R_{GMG}$ and $R_{GMMG}$ also decrease, by ~23% and ~27%, respectively. Taken together with the PL and Raman measurements, these results reveal a strong correlation between the strength of interlayer coupling and thermal resistance of the vdW junctions. Furthermore, the fact that $R_{GG}$ and $R_G$ are similar after annealing, and each is smaller than $R_{GMG}$, suggests that the intrinsic conductance of a *G-G* homojunction is significantly larger than that of a *G-M* heterojunction. Note that our ability to draw these conclusions is based on our confidence that the Al-*G* and *G*-SiO$_2$ interface resistances do not vary significantly between different regions on a sample since they are within ~100 μm of one another, and also do not change substantially upon annealing as $R_G$ is similar before and after the anneal. While we also measure the thermal resistance of *MG* (≈100 m$^2$K(GW)$^{-1}$), we do not draw any major conclusions from it since its top layer is MoS$_2$, whose thermal resistance with Al could be different from that of graphene.

To probe these effects in higher order stacks we examine sample D3, which is annealed. **Fig. 4A,B** present optical microscopy and TDTR thermal resistance maps of a region of the sample that has four stacking sequences: *GGG*, *GMGG*, *GGMG*, and *GMGMG*. A different region containing *GG* and *GMG* sequences is shown in **Fig. S9**. Single-spot delay scans (**Fig. 4C**) show that the temperature decay rates for stacks *GG* and *GGG* are very similar, as well as for stacks *GMG*, *GMGG*, and *GGMG*. Consistent with this, an ROI-analysis of TDTR maps (**Fig. 4E**) illustrates that the $R_{2D}$ distributions of *GG* and *GGG* overlap, as do those of *GMG*, *GMGG*, and *GGMG*. A summary of measurements made on this sample is presented in **Fig. 4D**. The interfacial thermal resistances follow the trend: $R_{GMGMG} > R_{GMGG} \approx R_{GGMG} \approx R_{GMG} > R_{GGG} \approx R_{GG}$.

We make three key observations from the data. First, the strongly coupled (i.e. annealed) *G*, *GG* and *GGG* interfaces have nearly equal thermal resistance, indicating that phonon transport across these interfaces is quasi-ballistic. This is shown in D3 by $R_{GGG} \approx R_{GG}$, and in D2 by $R_{GG} \approx R_G$, and is furthermore consistent with the observation that $R_{GMGG} \approx R_{GGMG} \approx R_{GMG}$ (**Fig. 4C-E**). We estimate the *effective* thermal resistance ($\rho$) of a single *G-G* junction, $\rho_{GG}$, using $(R_{GGG} - R_{GG})_{D3}$ and $(R_{GG} - R_G)_{D2}$, which gives $\rho_{GG} < 4$ m$^2$KGW$^{-1}$. This is consistent with prior calculations of the cross-plane ballistic thermal resistance of graphite at room temperature,[28] which gave ≈ 3 m$^2$KGW$^{-1}$. Similarly, we also extract the *effective* thermal resistance of an *M-M* junction, $\rho_{MM}$ using $(R_{GMMG} - R_{GMG})$ where $R_{GMG}$ is averaged between D2 and D3; this gives $\rho_{MM} \approx 26$ m$^2$KGW$^{-1}$. Our measurements of Al/*nG*/SiO$_2$ (*n* = 1, 2, 3) are in good agreement with previous data for the thermal resistance of exfoliated few-layer graphene in Au/Ti/*nG*/SiO$_2$ and Au/Ti/*nG*/Si heterostructures,[17,29] where no strong layer dependence was observed. This similarity between our LBL-assembled CVD-grown layers and prior results on pristine interfaces in exfoliated crystals suggests a good *G-G* interface quality in our annealed samples. Further, given that our polycrystalline samples are not prepared with well-defined interlayer twist, this also indicates that turbostratic disorder may not have a strong impact on *G-G* thermal coupling.

Second, we find that the order of stacking does not affect the total thermal resistance, as shown by $R_{GMGG} \approx R_{GGMG}$. This implies no measurable thermal rectification in such vdW junctions. This is consistent with prior MD simulations of graphene/MoS$_2$[30] and other 2D heterojunctions.[31] However, we do not exclude the possibility of rectification under significantly larger thermal gradients, when non-linearities in the vibrational spectra of *G* and *M* lattices may become important.[32] In the present experiments, the vertical temperature gradient is on the order of 1 K nm$^{-1}$.

Third, the heterojunction *G-M* is more resistive than the homojunctions *G-G* and *M-M*. We estimate the *effective G-M* resistance using $\rho_{GM} = (R_{GMGMG} - R_G)/4 \approx 37$ m$^2$KGW$^{-1}$. Comparing this with the thermal resistances of homojunctions, we summarize the trend as: $\rho_{GM} > \rho_{MM} \gg \rho_{GG}$. This trend is consistent with previous MD simulations[30] which had calculated $\rho_{GM} \approx 3\rho_{MM}$ and $\rho_{GM} \approx 15\ \rho_{GG}$. Notably, a single *G-M* heterojunction consisting of a graphene and MoS$_2$ monolayer placed only ~0.5 nm apart has a thermal resistance comparable to ~50 nm of SiO$_2$. Equivalently, the *GMGMG* stack with a total thickness of ~2 nm has an effective cross-plane thermal conductivity (= thickness/resistance) < 0.02 Wm$^{-1}$K$^{-1}$. This represents one of the lowest thermal conductivities among dense solids,[12,33] lower than that of dry air at 1 atm and 300 K, ~0.026 Wm$^{-1}$K$^{-1}$.[34]

To gain further insights into the mechanisms governing heat transport in vdW stacks, we performed NEMD simulations. The following stacks were simulated to compare with the experiments: *G*, *GG*, *GGG*, *GMG*, *GMMG*, and *GMGMG*. Each stack was enclosed within nine-atom thick Al leads. To account for thermal expansion at finite temperature, the structures were relaxed at 300 K using a free-energy minimization (FEM) approach in the quasi-harmonic approximation (QHA).[35–37] As we will show later, thermal expansion is crucial to accurately capture the thermal resistances of the vdW junctions, owing to the strong sensitivity of the TBR to the interlayer spacing. Details of the simulation approach are provided under Methods.

**Fig. 5A** shows the calculated thermal resistances of the various stacks (blue bars) plotted alongside the experimental data (red bars). In the calculations, the plotted results include the resistances of the top and bottom Al-*G* and *G*-Al interfaces, whereas in the experiments, they include the Al-*G* and *G*-SiO$_2$ interfaces. In general, good agreement is seen between the trends in the experiments and simulations, with a deviation of <20% for all stacks except *GMGMG*.

Notably, the simulations show that the resistances of the pure graphene junctions are independent of the number of layers, which is consistent with the measurements. This agreement is noteworthy considering that in the simulations adjacent *G* layers are AB stacked, whereas in the experiments the layers are arbitrarily twisted. This suggests that thermal transport across few-layer *G-G* junctions is quasi-ballistic and is not strongly sensitive to the interlayer atomic registry. This is also consistent with prior experiments on exfoliated AB-stacked few-layer graphene films where no layer-number dependence was observed in the cross-plane thermal resistance.[17] Next, the simulations predict the trend $R_{GMGMG} > R_{GMMG} > R_{GMG}$, which is consistent with the measurements. For *GMGMG*, two stacks are simulated, with AB and AA stacking of the *M* layers; the latter has a resistance that is ~8% larger than the former. The discrepancy between theory and experiments is the largest for *GMGMG*, at around 30%.

To understand possible reasons behind this, we consider the effect of interlayer spacing on thermal transport. To do this, we calculate the cross-plane thermal resistance of *GMMG* and *GMGMG* using NEMD, while varying the distance between two fixed layers of Al atoms within the top and bottom leads. In **Fig. 5B**, we plot $R_{2D}$ versus the percentage cross-plane lattice expansion calculated relative to its value at $T = 0$ K (purple and green curves). We see a strong, non-linear increase in resistance with interlayer spacing: for instance, an expansion of only 4% leads to an increase of ~400% in the thermal resistance of *GMGMG*. This is qualitatively consistent with prior NEMD simulations[38] and *ab initio* lattice dynamics calculations[39] of the effect of cross-plane tensile strain on the c-axis thermal conductivity of pure $MoS_2$. In **Fig. 5B**, we also plot points corresponding to the expansion and $R_{2D}$ values at 300 K using the FEM approach, which are in good agreement with the 0 K expansion curves. Based on these calculations, we suggest that the average interlayer spacing in the *GMGMG* sample is ~1% larger than the equilibrium spacing at 300 K, which causes NEMD to underpredict the measured resistance by ~30%.

We draw two conclusions from this analysis. First, as the number of layers in the experimental vdW stack increases, the *average* interlayer separation exceeds the equilibrium value, possibly owing to trapped contaminants. Second, the strong sensitivity of the thermal resistance to c-axis lattice expansion indicates that extreme care must be taken to ensure cleanliness of heterostructures, to be able to make fundamental measurements of interlayer phonon transport.

In conclusion, using a combinatorial experimental approach we have systematically characterized cross-plane thermal transport in LBL-assembled van der Waals stacks made of graphene and $MoS_2$. Using correlative time-domain thermoreflectance and photoluminescence spectroscopy, in conjunction with molecular dynamics simulations, we have examined the effects of vibrational mismatch, junction asymmetry, and interlayer coupling strength on heat transport across vdW interfaces. Our results provide a framework for understanding (tunable) heat flow in a broad range of vdW metamaterials and enable the creation of an artificial dense solid with an effective thermal conductivity lower than that of air. Lastly, the insights presented here will allow for improved engineering of heat flow at vdW heterojunctions in 2D electronics, which is crucial for achieving ultimate performance limits in emerging computing and photonic devices.

**Methods**

CVD growth of graphene and $MoS_2$: Continuous monolayer graphene samples were prepared *via* a Cu-mediated low-pressure CVD process. Cu foil (≥ 99.9 % purity, 33 μm thick, JX Nippon Mining & Metals) was placed in a 2-inch quartz tube after cleaning with acetic acid. After annealing the Cu foil under hydrogen atmosphere at 1040°C, graphene was synthesized by flowing methane and hydrogen gas at 1040°C for 40 minutes. Discontinuous monolayer $MoS_2$ samples consisting of triangular-shaped crystals were prepared *via* an atmospheric-pressure CVD process on a $SiO_2$/Si substrate. The $SiO_2$/Si substrate was first decorated with 20 μl of a perylene-3,4,9,10 tetracarboxylic acid tetrapotassium (PTAS) solution. It was placed in the center of a 2-inch quartz tube facing ~0.6 mg of $MoO_3$, while sulfur was located upstream. $MoS_2$ was synthesized by the interaction between the sulfur carried by 2 sccm of argon inflow and the vaporized $MoO_3$ on the substrate. To obtain discontinuous monolayers, the growth temperature and time were carefully tailored and adjusted in the range of 700-800°C and 10-20 minutes, respectively, and the position of sulfur was optimized. The samples were prepared without controlling the interlayer twist angles.

Polymer-assisted transfer process: Multilayer stacks comprising graphene (*G*) and $MoS_2$ (*M*) monolayers were prepared *via* multiple polymer-assisted layer-by-layer transfer processes. An as-grown graphene film on Cu foil was spin-coated with Poly(methyl methacrylate) (PMMA, 950K A4) and Polystyrene (PS, 2.8 g/mol). The Cu foil was etched away by placing the sample in a Cu etchant (CE-100, Transene) for 1 hour. The resulting *PS/PMMA/G* stack was rinsed thoroughly in a deionized water bath. The stack was pinched out of the bath and placed onto an as grown $MoS_2$ sample on a $SiO_2$/Si substrate. Note that the *G/M* interface was dry i.e. it did not see polymer. While maintaining adhesion between graphene and $MoS_2$, a droplet of water was used to delaminate $MoS_2$ from the $SiO_2$/Si substrate,[40] resulting in a *PS/PMMA/G/M* stacked sample. Higher order stacks were prepared by iterating this procedure as required. For example, to assemble the *GMGMG* stack, the next step involved transferring this stack onto another as grown graphene sample on Cu foil and repeating the above steps. Note again that the transfer process did not introduce polymer between the 2D layers. The final step involved transferring this stack (*e.g. PS/PMMA/G/M/G/M/G*) onto a $SiO_2$ (90 nm) on Si substrate. The *PS/PMMA* layers were removed by soaking the sample in toluene.

Time-domain thermoreflectance: TDTR is an optical pump-probe technique, which is used to measure thermal conductivity of thin films and TBR of interfaces. Details of this method and our setup are described elsewhere.[2,7,19] Briefly, our setup is based on a 1064 nm, 82 MHz oscillator that produces ~9 ps pulses. Pump pulses are frequency-doubled to 532 nm by second harmonic generation, and amplitude modulated at a frequency of $f_{mod}$ = 10 MHz using an electro-optic modulator. An optically opaque 80 nm thick Al transducer layer absorbs the pulses and converts them to heat. As the heat pulses diffuse through the film of interest, in our case the stack of 2D layers, the temperature decay of the transducer is monitored by measuring the reflected probe intensity as a function of delay time between the pump and probe (0 to 3.7 ns). The measured data consists of the in-phase and out-of-phase voltage components of the probe intensity demodulated at $f_{mod}$ using a lock-in amplifier, $V_{in}$ and $V_{out}$, respectively. In a typical measurement, the time-series of voltage ratio data (= $-V_{in}/V_{out}$) is fit to the solution of a 3D heat diffusion model to extract $R_{2D}$.

TDTR mapping: In the present experiments, we adapted the TDTR technique to create maps of $R_{2D}$ by fixing the pump-probe delay time at +250 ps and raster scanning the sample. At each position on the sample, $R_{2D}$ was extracted by comparing the voltage ratio (= $-V_{in}/V_{out}$) with a fixed-delay correlation curve obtained from the thermal model (**Fig. S1**). Spatial variations in the interfacial resistance due to different layer stacking appear predominantly as variations in $V_{out}$ (**Fig. S2**). The thickness and thermophysical properties of Al and $SiO_2$ layers, and Si substrate were either measured or taken from literature. In the current experiments, we used a root-mean-square laser spot size of ≈5.1 µm, and pump and probe incident powers of approximately 12 and 3 mW, respectively. The estimated steady state temperature rise is ~2 K. The FWHM spatial resolution of this technique is ≈2.2 µm (**Fig. S5**). We and others have previously used this technique to measure spatially-inhomogeneous thermal conductivity in polycrystalline diamond,[19] lithium-intercalated $MoS_2$,[2] and other materials.[14,29,41,42]

Molecular dynamics simulations:

Interatomic interactions are modeled with empirical potentials tested to reproduce the vibrational properties of graphite, $MoS_2$, and aluminum: specifically the reparametrized Kolmogorov-Crespi (KC)[43] was used for the interlayer graphene-graphene interactions, Lennard-Jones (LJ)[44] for the interlayer graphene-$MoS_2$ interactions, optimized Tersoff[45] for the in-plane graphene interactions, reactive empirical bond-order (REBO)-type Mo-S potential[46] for all $MoS_2$-$MoS_2$ interactions, Morse potential[47] for the Al-graphene interactions, and embedded atom model (EAM)[48] for the Al-Al interactions.

The models were built replicating the commensurate 4:5 $MoS_2$-graphene hexagonal unit cell, which is made of 4x4 $MoS_2$ unit cells and 5x5 graphene unit cells.[49] The resulting *GMG* cells were then made commensurate with aluminum by creating a 13:3 Al-*GMG* unit cell using an aluminum *fcc* cell with a lattice parameter of 0.286 nm. (111) Al slabs were placed on both sides of the *GMG* heterojunctions. We adopt periodic boundary conditions in the plane perpendicular to transport and fixed boundary conditions in the transport direction. The latter were implemented by constraining the coordinates of the outermost layer of Al atoms. We took special care to optimize the interlayer separation in the device by minimizing the free energy with the quasi-harmonic approximation[35–37] (QHA) instead of classical MD in the constant pressure canonical ensemble, as the TBR is critically sensitive to cross-plane expansion as shown by **Fig. 5B**. We minimize the free energy with the QHA because it yields a 0→300 K graphite thermal expansion of 1.12%, which is in better agreement with the measured thermal expansion[50] of 1.54% than MD simulations with the Nosé-Hoover barostat, which yields an expansion of 0.53%. The better agreement between QHA thermal expansion and experiment over MD likely results from the inclusion of quantum statistics in the QHA, which help to correctly include the populations of soft flexural modes. The QHA free energy was minimized with respect to c-axis expansion and was computed by $F(V,T) = E_{0K}(V) + F_{vib}(V,T)$ where $E_{0K}$ is the total energy of the system at 0 K at a given volume, and $F_{vib}$ represents the vibrational contribution to the free energy: $\sum_\nu \frac{\hbar\omega_\nu}{2} + \frac{\log(1-e^{-\beta\hbar\omega_\nu})}{\beta}$ where $\omega_\nu^2$ are the dynamical matrix eigenvalues yielded from all the atoms in the system except for the outer three layers of Al slabs on both sides of the device, and $\beta = (k_B T)^{-1}$. The dynamical matrices used for calculating the vibrational free energy were computed with LAMMPS[51] using the finite difference method with an atomic displacement of $10^{-7}$ nm. Once the interlayer separation was minimized according to the free energy, we calculated the thermal resistance of the devices by NEMD, following the protocol established in Ref.[52] An outer layer of Al was fixed while the middle three layers of Al (out of nine layers) on each side were thermostatted to 350 K and 250 K with a Langevin integrator with a 1 ps relaxation rate. The equations of motion were integrated with a 1 fs timestep. NEMD simulations were run, using LAMMPS,[51] for 4.5 ns to ensure that a steady state heat flux was reached. Simulation configurations and temperature profiles computed in these simulations are shown in **Fig. S10** and **Fig. S11**.

**Author contributions**

A.S., Y.C.S. and E.P. conceived the research; A.S. performed the measurements and analyzed all experimental data; Y.C.S. fabricated the samples with assistance from V.C. and K.K.H.S.; C.S. performed the molecular dynamics simulations with inputs from S.Che. and D.D.; S.Cha. assisted with TEM characterization.; A.S. wrote the manuscript with inputs from C.S., Y.C.S., D.D. and E.P.; D.D., K.E.G. and E.P. supervised the project.

**Acknowledgements**

Part of this work was performed at the Stanford Nano Shared Facilities (SNSF)/Stanford Nanofabrication Facility (SNF), supported by the National Science Foundation under award ECCS-2026822. This research was supported in part by the NSF Engineering Research Center for Power Optimization of Electro Thermal Systems (POETS) with cooperative agreement EEC-1449548, by AFOSR Grant FA9550-14-1-0251, by NSF EFRI 2-DARE Grant 1542883, and by the Stanford SystemX Alliance. We also acknowledge Brookhaven National Lab for allocating computational resources. This research used resources of the Center for Functional Nanomaterials, which is a U.S. DOE Office of Science Facility, at Brookhaven National Laboratory under Contract No. DE-SC0012704.

## References

(1) Cho, J.; Losego, M. D.; Zhang, H. G.; Kim, H.; Zuo, J.; Petrov, I.; Cahill, D. G.; Braun, P. V. Electrochemically Tunable Thermal Conductivity of Lithium Cobalt Oxide. *Nat. Commun.* **2014**, *5*, 4035. https://doi.org/10.1038/ncomms5035.

(2) Sood, A.; Xiong, F.; Chen, S.; Wang, H.; Selli, D.; Zhang, J.; McClellan, C. J.; Sun, J.; Donadio, D.; Cui, Y.; et al. An Electrochemical Thermal Transistor. *Nat. Commun.* **2018**, *9*, 4510. https://doi.org/10.1038/s41467-018-06760-7.

(3) Li, D.; Wu, Y.; Kim, P.; Shi, L.; Yang, P.; Majumdar, A. Thermal Conductivity of Individual Silicon Nanowires. *Appl. Phys. Lett.* **2003**, *83* (14), 2934–2936. https://doi.org/10.1063/1.1616981.

(4) Geim, A. K.; Grigorieva, I. V. Van Der Waals Heterostructures. *Nature* **2013**, *499*, 419–425. https://doi.org/10.1038/nature12385.

(5) Guo, R.; Jho, Y. D.; Minnich, A. J. Coherent Control of Thermal Phonon Transport in van Der Waals Superlattices. *Nanoscale* **2018**, *10*, 14432–14440. https://doi.org/10.1039/c8nr02150c.

(6) Hu, S.; Ju, S.; Shao, C.; Guo, J.; Xu, B.; Ohnishi, M.; Shiomi, J. Ultimate Impedance of Coherent Heat Conduction in van Der Waals Graphene-MoS2 Heterostructures. *Mater. Today Phys.* **2020**, *16*, 100324. https://doi.org/10.1016/j.mtphys.2020.100324.

(7) Sood, A.; Xiong, F.; Chen, S.; Cheaito, R.; Lian, F.; Asheghi, M.; Cui, Y.; Donadio, D.; Goodson, K. E.; Pop, E. Quasi-Ballistic Thermal Transport Across MoS2 Thin Films. *Nano Lett.* **2019**, *19* (4), 2434–2442. https://doi.org/10.1021/acs.nanolett.8b05174.

(8) Sadeghi, H.; Sangtarash, S.; Lambert, C. J. Cross-Plane Enhanced Thermoelectricity and Phonon Suppression in Graphene/MoS2 van Der Waals Heterostructures. *2D Mater.* **2017**, *4*, 015012. https://doi.org/10.1088/2053-1583/4/1/015012.

(9) Suryavanshi, S. V.; Gabourie, A. J.; Barati Farimani, A.; Pop, E. Thermal Boundary Conductance of Two-Dimensional $MoS_2$ Interfaces. *J. Appl. Phys.* **2019**, *126* (5), 055107. https://doi.org/10.1063/1.5092287.

(10) Zhao, Y.; Cai, Y.; Zhang, L.; Li, B.; Zhang, G.; Thong, J. T. L. Thermal Transport in 2D Semiconductors — Considerations for Device Applications. *Adv. Funct. Mater.* **2020**, *30*, 1903929. https://doi.org/10.1002/adfm.201903929.

(11) Yalon, E.; Mcclellan, C. J.; Smithe, K. K. H.; Xu, R. L.; Rojo, M. M.; Suryavanshi, S. V.; Gabourie, A. J.; Neumann, C. M.; Xiong, F.; Pop, E. Energy Dissipation in Monolayer MoS2 Electronics. *Nano Lett.* **2017**, *17* (6), 3429–3433. https://doi.org/10.1021/acs.nanolett.7b00252.

(12) Vaziri, S.; Yalon, E.; Rojo, M. M.; Suryavanshi, S. V; Zhang, H.; Mcclellan, C. J.; Bailey, C. S.; Smithe, K. K. H.; Gabourie, A. J.; Chen, V.; et al. Ultrahigh Thermal Isolation across Heterogeneously Layered Two-Dimensional Materials. *Sci. Adv.* **2019**, *5*, eaax1325. https://doi.org/10.1126/sciadv.aax1325.

(13) Liu, Y.; Ong, Z.; Wu, J.; Zhao, Y.; Watanabe, K.; Taniguchi, T.; Chi, D.; Zhang, G.; Thong, J. T. L.; Qiu, C.-W.; et al. Thermal Conductance of the 2D MoS2/h-BN and Graphene/h-BN Interfaces. *Sci. Rep.* **2017**, *7*, 43886. https://doi.org/10.1038/srep43886.

(14) Brown, D. B.; Shen, W.; Li, X.; Xiao, K.; Geohegan, D. B.; Kumar, S. Spatial Mapping of Thermal Boundary Conductance at Metal-Molybdenum Diselenide Interfaces. *ACS Appl. Mater. Interfaces* **2019**, *11* (15), 14418–14426. https://doi.org/10.1021/acsami.8b22702.

(15) Yasaei, P.; Foss, C. J.; Karis, K.; Behranginia, A.; El-Ghandour, A. I.; Fathizadeh, A.; Olivares, J.; Majee, A. K.; Foster, C. D.; Khalili-Araghi, F.; et al. Interfacial Thermal Transport in Monolayer MoS2- and Graphene-Based Devices. *Adv. Mater. Interfaces* **2017**, *4*, 1700334. https://doi.org/10.1002/admi.201700334.

(16) Hopkins, P. E.; Baraket, M.; Barnat, E. V.; Beechem, T. E.; Kearney, S. P.; Duda, J. C.; Robinson, J. T.; Walton, S. G. Manipulating Thermal Conductance at Metal-Graphene Contacts via Chemical Functionalization. *Nano Lett.* **2012**, *12* (2), 590–595. https://doi.org/10.1021/nl203060j.

(17) Koh, Y. K.; Bae, M. H.; Cahill, D. G.; Pop, E. Heat Conduction across Monolayer and Few-Layer Graphenes. *Nano Lett.* **2010**, *10* (11), 4363–4368. https://doi.org/10.1021/nl101790k.

(18) Nyby, C.; Sood, A.; Zalden, P.; Gabourie, A. J.; Muscher, P.; Rhodes, D.; Mannebach, E.; Corbett, J.; Mehta, A.; Pop, E.; et al. Visualizing Energy Transfer at Buried Interfaces in Layered Materials Using Picosecond X-Rays. *Adv. Funct. Mater.* **2020**, *30*, 2002282. https://doi.org/10.1002/adfm.202002282.

(19) Sood, A.; Cheaito, R.; Bai, T.; Kwon, H.; Wang, Y.; Li, C.; Yates, L.; Bougher, T.; Graham, S.; Asheghi, M.; et al. Direct Visualization of Thermal Conductivity Suppression Due to Enhanced Phonon Scattering near Individual Grain Boundaries. *Nano Lett.* **2018**, *18* (6), 3466–3472. https://doi.org/10.1021/acs.nanolett.8b00534.

(20) Smithe, K. K. H.; English, C. D.; Suryavanshi, S. V.; Pop, E. Intrinsic Electrical Transport and Performance Projections of Synthetic Monolayer MoS2 Devices. *2D Mater.* **2017**, *4*, 011009. https://doi.org/10.1088/2053-1583/4/1/011009.

(21) Wang, N. C.; Carrion, E. A.; Tung, M. C.; Pop, E. Reducing Graphene Device Variability with Yttrium Sacrificial Layers. *Appl. Phys. Lett.* **2017**, *110*, 223106. https://doi.org/10.1063/1.4984090.

(22) Zhu, G.; Liu, J.; Zheng, Q.; Zhang, R.; Li, D.; Banerjee, D.; Cahill, D. G. Tuning Thermal Conductivity in Molybdenum Disulfide by Electrochemical Intercalation. *Nat. Commun.* **2016**, *7*, 13211. https://doi.org/10.1038/ncomms13211.

(23) Mak, K. F.; Lee, C.; Hone, J.; Shan, J.; Heinz, T. F. Atomically Thin MoS2: A New Direct-Gap Semiconductor. *Phys. Rev. Lett.* **2010**, *105*, 136805. https://doi.org/10.1103/PhysRevLett.105.136805.

(24) Liu, K.; Zhang, L.; Cao, T.; Jin, C.; Qiu, D.; Zhou, Q.; Zettl, A.; Yang, P.; Louie, S. G.; Wang, F. Evolution of Interlayer Coupling in Twisted Molybdenum Disulfide Bilayers. *Nat. Commun.* **2014**, *5*, 4966. https://doi.org/10.1038/ncomms5966.

(25) Hemmat, Z.; Yasaei, P.; Schultz, J. F.; Hong, L.; Majidi, L.; Behranginia, A.; Verger, L.; Jiang, N.; Barsoum, M. W.; Klie, R. F.; et al. Tuning Thermal Transport Through Atomically Thin Ti3C2Tz MXene by Current Annealing in Vacuum. *Adv. Funct. Mater.* **2019**, *29*, 1805693. https://doi.org/10.1002/adfm.201805693.

(26) Jin, K.; Liu, D.; Tian, Y. Enhancing the Interlayer Adhesive Force in Twisted Multilayer MoS2 by Thermal Annealing Treatment. *Nanotechnology* **2015**, *26* (40), 405708. https://doi.org/10.1088/0957-4484/26/40/405708.

(27) Das, A.; Chakraborty, B.; Piscanec, S.; Pisana, S.; Sood, A. K.; Ferrari, A. C. Phonon Renormalization in Doped Bilayer Graphene. *Phys. Rev. B* **2009**, *79*, 155417. https://doi.org/10.1103/PhysRevB.79.155417.

(28) Li, Z.; Liu, Y.; Lindsay, L.; Xu, Y.; Duan, W.; Pop, E. Size Dependence and Ballistic Limits of Thermal Transport in Anisotropic Layered Two-Dimensional Materials. *arXiv:1711.02772* **2017**.

(29) Yang, J.; Ziade, E.; Maragliano, C.; Crowder, R.; Wang, X.; Stefancich, M.; Chiesa, M.; Swan, A. K.; Schmidt, A. J. Thermal Conductance Imaging of Graphene Contacts. *J. Appl. Phys.* **2014**, *116*, 023515. https://doi.org/10.1063/1.4889928.

(30) Ding, Z.; Pei, Q. X.; Jiang, J. W.; Huang, W.; Zhang, Y. W. Interfacial Thermal Conductance in Graphene/MoS2 Heterostructures. *Carbon* **2016**, *96*, 888–896. https://doi.org/10.1016/j.carbon.2015.10.046.

(31) Zhang, J.; Hong, Y.; Yue, Y. Thermal Transport across Graphene and Single Layer Hexagonal Boron Nitride. *J. Appl. Phys.* **2015**, *117* (13), 134307. https://doi.org/10.1063/1.4916985.

(32) Li, B.; Wang, L.; Casati, G. Thermal Diode: Rectification of Heat Flux. *Phys. Rev. Lett.* **2004**, *93* (18), 184301. https://doi.org/10.1103/PhysRevLett.93.184301.

(33) Chiritescu, C.; Cahill, D. G.; Nguyen, N.; Johnson, D.; Bodapati, A.; Keblinski, P.; Zschack, P. Ultralow Thermal Conductivity in Disordered, Layered WSe$_2$ Crystals. *Science* **2007**, *315* (5810), 351–353. https://doi.org/10.1126/science.1136494.

(34) Kadoya, K.; Matsunaga, N.; Nagashima, A. Viscosity and Thermal Conductivity of Dry Air in the Gaseous Phase. *J. Phys. Chem. Ref. Data* **1985**, *14* (4), 947–970. https://doi.org/10.1063/1.555744.

(35) Leibfried, G.; Ludwig, W. Theory of Anharmonic Effects in Crystals. *Solid State Physics*; Seitz,

F., Turnbull, D., Eds.; Academic Press: New York, 1961; pp 276–444.

(36) Errea, I.; Calandra, M.; Mauri, F. Anharmonic Free Energies and Phonon Dispersions from the Stochastic Self-Consistent Harmonic Approximation: Application to Platinum and Palladium Hydrides. *Phys. Rev. B* **2014**, *89*, 064302. https://doi.org/10.1103/PhysRevB.89.064302.

(37) Nath, P.; Plata, J. J.; Usanmaz, D.; Al Rahal Al Orabi, R.; Fornari, M.; Nardelli, M. B.; Toher, C.; Curtarolo, S. High-Throughput Prediction of Finite-Temperature Properties Using the Quasi-Harmonic Approximation. *Comput. Mater. Sci.* **2016**, *125*, 82–91. https://doi.org/10.1016/j.commatsci.2016.07.043.

(38) Ding, Z.; Jiang, J.-W.; Pei, Q.-X.; Zhang, Y.-W. In-Plane and Cross-Plane Thermal Conductivities of Molybdenum Disulfide. *Nanotechnology* **2015**, *26* (6), 65703. https://doi.org/10.1088/0957-4484/26/6/065703.

(39) Chen, S.; Sood, A.; Pop, E.; Goodson, K. E.; Donadio, D. Strongly Tunable Anisotropic Thermal Transport in MoS2 by Strain and Lithium Intercalation: First-Principles Calculations. *2D Mater.* **2019**, *6* (2), 025033. https://doi.org/10.1088/2053-1583/ab0715.

(40) Gurarslan, A.; Yu, Y.; Su, L.; Yu, Y.; Suarez, F.; Yao, S.; Zhu, Y.; Ozturk, M.; Zhang, Y.; Cao, L. Surface-Energy-Assisted Perfect Transfer of Centimeter-Scale Monolayer and Few-Layer MoS2 Films onto Arbitrary Substrates. *ACS Nano* **2014**, *8* (11), 11522–11528. https://doi.org/10.1021/nn5057673.

(41) Huxtable, S.; Cahill, D. G.; Fauconnier, V.; White, J. O.; Zhao, J. C. Thermal Conductivity Imaging at Micrometre-Scale Resolution for Combinatorial Studies of Materials. *Nat. Mater.* **2004**, *3*, 298–301. https://doi.org/10.1038/nmat1114.

(42) Olson, D. H.; Avincola, V. A.; Parker, C. G.; Braun, J. L.; Gaskins, J. T.; Tomko, J. A.; Opila, E. J.; Hopkins, P. E. Anisotropic Thermal Conductivity Tensor of β-Y2Si2O7 for Orientational Control of Heat Flow on Micrometer Scales. *Acta Mater.* **2020**, *189*, 299–305. https://doi.org/10.1016/j.actamat.2020.02.040.

(43) Kolmogorov, A. N.; Crespi, V. H. Registry-Dependent Interlayer Potential for Graphitic Systems. *Phys. Rev. B* **2005**, *71*, 235415. https://doi.org/10.1103/PhysRevB.71.235415.

(44) Jiang, J.-W.; Park, H. S. Mechanical Properties of MoS2/Graphene Heterostructures. *Appl. Phys. Lett.* **2014**, *105*, 033108. https://doi.org/10.1063/1.4891342.

(45) Lindsay, L.; Broido, D. Optimized Tersoff and Brenner Empirical Potential Parameters for Lattice Dynamics and Phonon Thermal Transport in Carbon Nanotubes and Graphene. *Phys. Rev. B* **2010**, *81*, 205441. https://doi.org/10.1103/PhysRevB.81.205441.

(46) Liang, T.; Phillpot, S. R.; Sinnott, S. B. Parametrization of a Reactive Many-Body Potential for Mo – S Systems. *Phys. Rev. B* **2009**, *79*, 245110. https://doi.org/10.1103/PhysRevB.79.245110.

(47) Reshetniak, V. V.; Aborkin, A. V. Aluminum–Carbon Interaction at the Aluminum–Graphene and Aluminum–Graphite Interfaces. *J. Exp. Theor. Phys.* **2020**, *130* (2), 214–227. https://doi.org/10.1134/S1063776120010173.

(48) Mishin, Y.; Farkas, D.; Mehl, M. J.; Papaconstantopoulos, D. A. Interatomic Potentials for Monoatomic Metals from Experimental Data and Ab Initio Calculations. *Phys. Rev. B* **1999**, *59* (5), 3393–3407. https://doi.org/10.1103/PhysRevB.59.3393.

(49) Ebnonnasir, A.; Narayanan, B.; Kodambaka, S.; Ciobanu, C. Tunable MoS2 Bandgap in MoS2-Graphene Heterostructures. *Appl. Phys. Lett.* **2014**, *105*, 031603. https://doi.org/10.1063/1.4891430.

(50) Baskin, Y.; Meyer, L. Lattice Constants of Graphite at Low Temperatures. *Phys. Rev.* **1955**, *100* (2), 544. https://doi.org/10.1103/PhysRev.100.544.

(51) Plimpton, S. Fast Parallel Algorithms for Short-Range Molecular Dynamics. *J. Comput. Phys.* **1995**, *117*, 1–19. https://doi.org/10.1006/jcph.1995.1039.

(52) Li, Z.; Xiong, S.; Sievers, C.; Hu, Y.; Fan, Z.; Wei, N.; Bao, H.; Chen, S.; Donadio, D.; Ala-Nissila, T. Influence of Thermostatting on Nonequilibrium Molecular Dynamics Simulations of Heat Conduction in Solids. *J. Chem. Phys.* **2019**, *151*, 234105. https://doi.org/10.1063/1.5132543.

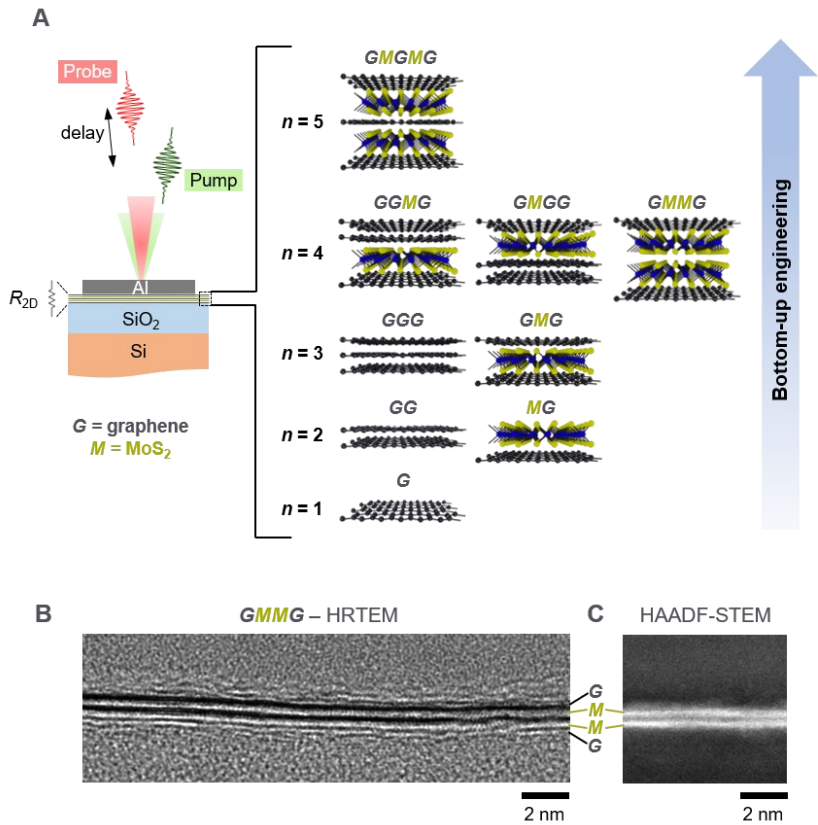

**Figure 1**. Combinatorial stacking of atomically-thin layers. (A) Cross-sectional schematic of the sample, comprising Al (80 nm)/{2D stack}/SiO$_2$ (90 nm)/Si substrate, where the '2D stack' can be one of nine sequences as shown on the right: *G*, *GG*, *MG*, *GGG*, *GMG*, *GGMG*, *GMGG*, *GMMG*, and *GMGMG*, with *n* denoting the number of layers. (B) High-resolution transmission electron microscope (HRTEM) and (C) High-angle annular dark-field scanning transmission electron microscope (HAADF-STEM) images of a cross-section of *GMMG*.

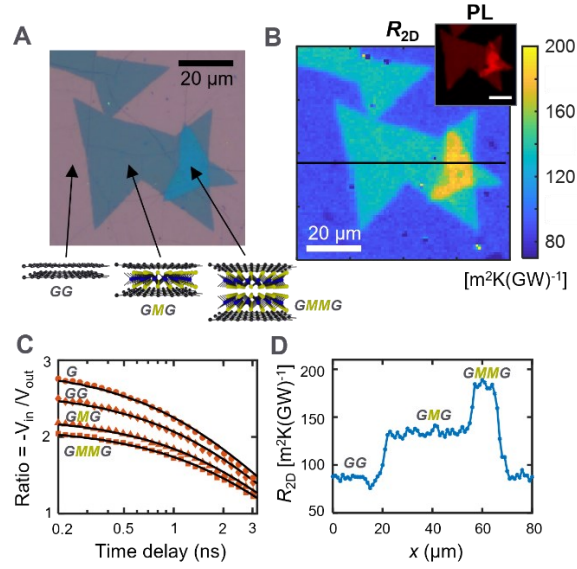

**Figure 2**. Correlative thermal, optical, and spectroscopic imaging. (A) Optical micrograph of sample D1 with stacks up to *GMMG*. Three regions are labeled, *GG*, *GMG* and *GMMG*. This sample was not annealed. (B) Map of the TBR between Al and $SiO_2$ ($R_{2D}$) measured by TDTR microscopy. Inset: $MoS_2$ PL map showing brighter signal from *GMMG* compared to *GMG* due to weak *M-M* coupling (scale bar 20 μm). A close correspondence is observed between the TDTR, optical, and PL micrographs. (C) TDTR time delay scans showing decay rates decreasing in the order *G > GG > GMG > GMMG*, corresponding to $R_{2D}$ increasing with layer number. (D) Line scan of $R_{2D}$ along the solid black line in (B).

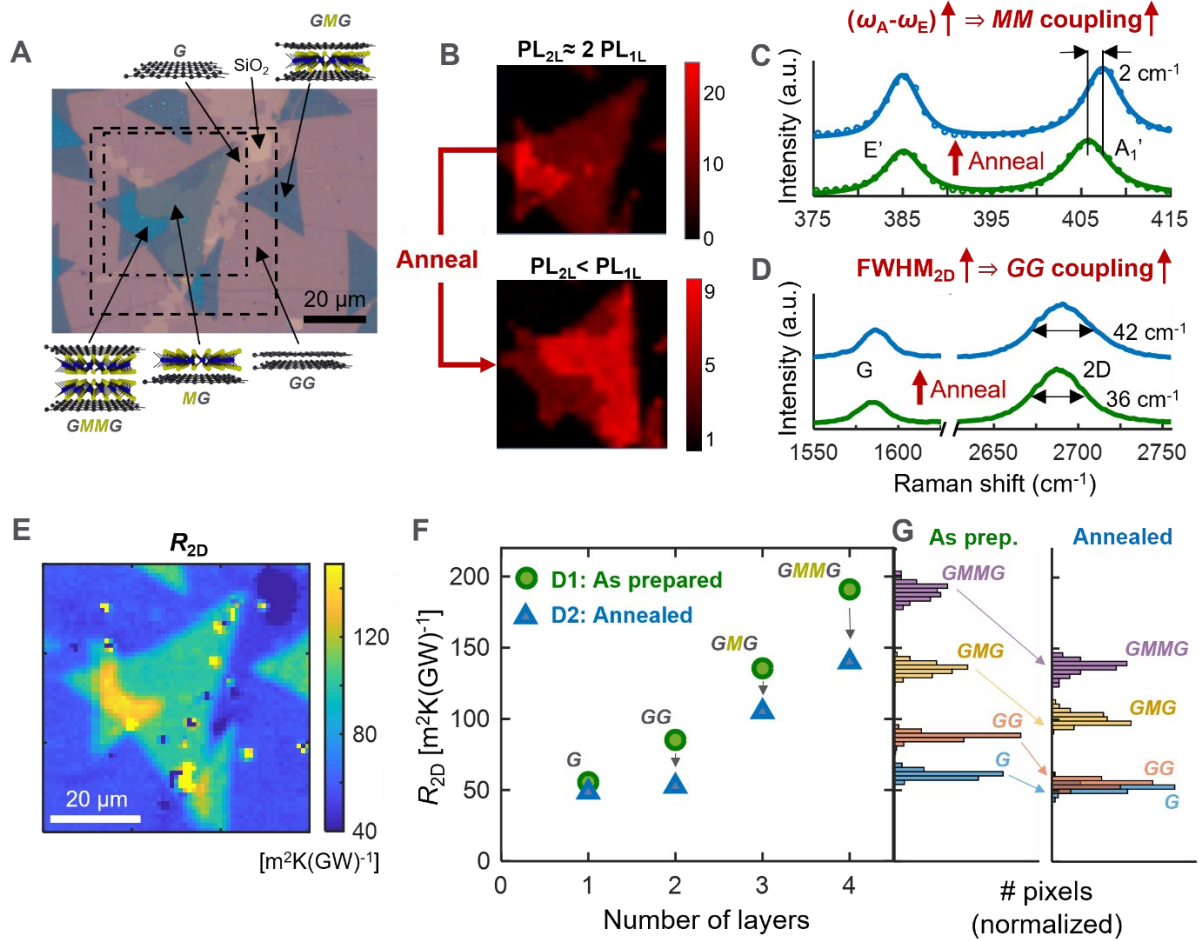

**Figure 3.** Vacuum annealing tunes interlayer coupling and thermal transport. (A) Optical micrograph of annealed sample D2 showing stacking sequences *G*, *GG*, *MG*, *GMG* and *GMMG*. (B) MoS$_2$ PL maps of the region marked by the dash-dotted lines in (A) before and after annealing at 350°C for 3 hours under vacuum. A crossover is observed after annealing, with the PL of *GMMG* becoming quenched relative to *GMG*, indicating stronger *M-M* coupling. Note that the sample is rotated by ~20° in the bottom panel; also, the image is slightly distorted, possibly due to stage drift. (C) MoS$_2$ Raman spectra of *GMMG* before (green) and after (blue) the anneal. A larger frequency difference between the A$_1$' and E' modes suggests enhanced *M-M* vibrational coupling upon annealing. (D) Graphene Raman spectra of *GG* before (green) and after (blue) the anneal. An increase in the width of the 2D peak, and a decrease in the intensity ratio of 2D and G peaks indicates strengthened *G-G* coupling. (E) TDTR map of $R_{2D}$ for the region marked by the dashed lines in (A). (F) $R_{2D}$ of stacks *G*, *GG*, *GMG*, and *GMMG* in samples D1 (as prepared, i.e. not annealed) and D2 (annealed), extracted from single-spot time scans. Error bars are omitted for clarity (see **Table S1** for uncertainty analysis). (G) Histograms of $R_{2D}$ for various stacking sequences based on regions of interest shown in **Fig. S7** and **Fig. S8**.

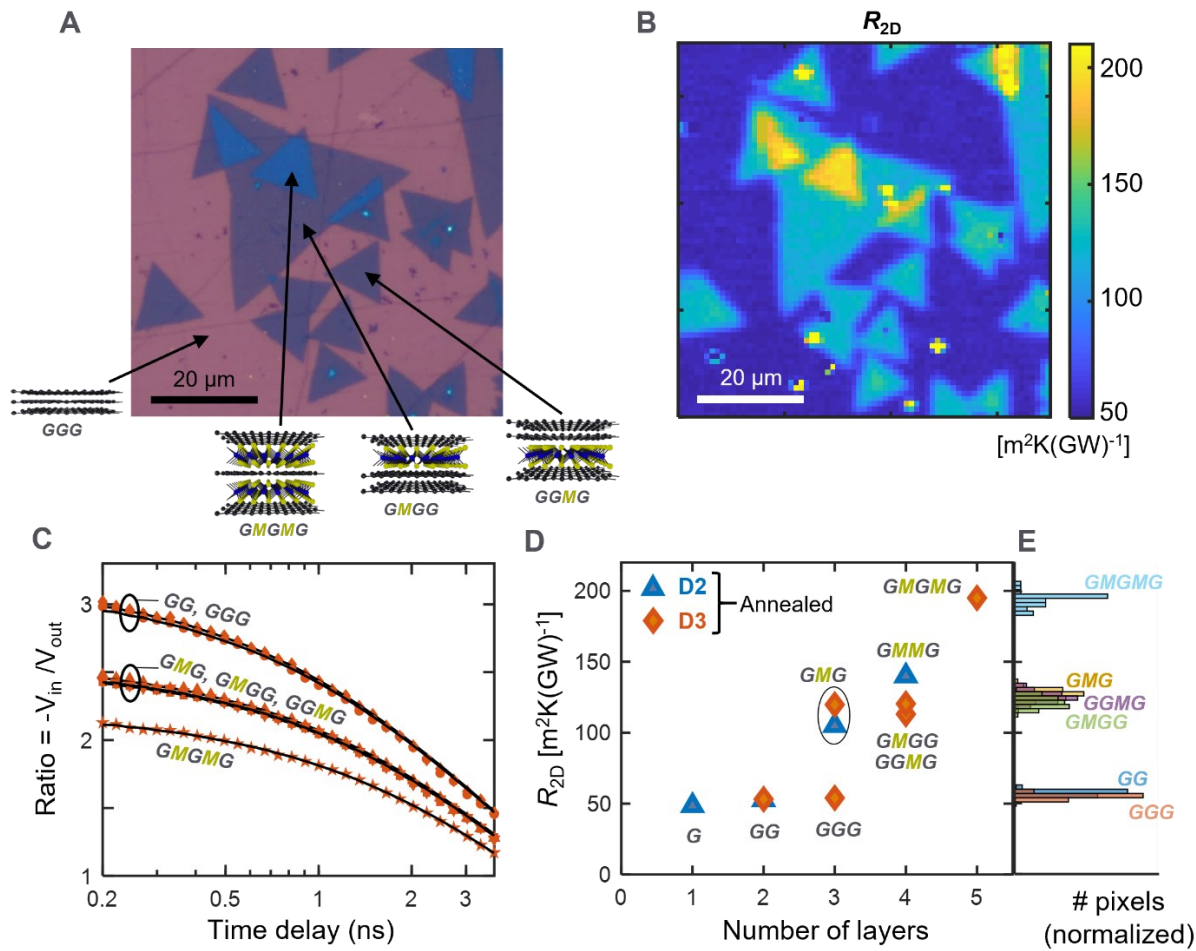

**Figure 4**. Towards ultralow thermal conductivity in higher order heterostructures. (A) Optical micrograph of sample D3 showing stacking sequences *GGG*, *GMGG*, *GGMG* and *GMGMG*. (B) TDTR map of $R_{2D}$; note that this sample is annealed. (C) Single-spot TDTR time scans. (D) $R_{2D}$ of all sequences combining data from samples D2 and D3, as extracted from single-spot measurements. Error bars are omitted for clarity (see **Table S1**). (E) Histograms of $R_{2D}$ for the different stacks in sample D3, based on regions of interest defined in **Fig. S9**. (C)-(E) show that the thermal resistances of *GGMG* and *GMGG* stacks are nearly identical, implying no thermal rectification. The highest order *GMGMG* stack has a thermal resistance that is equivalent to nearly 200 nm of $SiO_2$ even though it is 100× thinner.

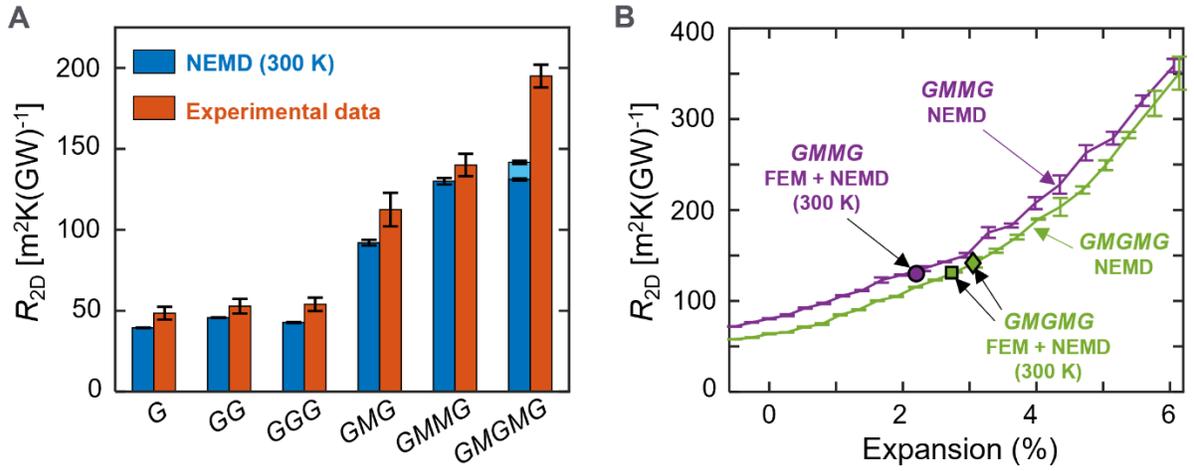

**Figure 5**. Molecular dynamics simulations and the effect of interlayer spacing on thermal transport. (A) NEMD calculations (blue bars) and experimental data (red bars) for various stacks. NEMD calculations are performed after relaxing each of the structures at 300 K, using a free-energy minimization approach that accounts for finite-temperature effects on the interlayer spacing. For *GMGMG*, the light blue bar denotes additional resistance when the *MM* layers are AA *vs*. AB stacked. For both theory and experiment, the plotted resistance includes the top and bottom interfaces with the leads. (B) NEMD-calculated thermal resistance of *GMMG* (purple curve) and *GMGMG* (green curve) as a function of cross-plane lattice expansion relative to zero Kelvin. Filled markers correspond to stacks with free-energy minimized lattice spacings at 300 K (purple circle—*GMMG*, green square—*GMGMG* with AB stacking between *M* layers, green diamond—*GMGMG* with AA stacking between *M* layers).

# Supplementary Materials for:

# Engineering thermal transport across layered graphene-MoS$_2$ superlattices

Aditya Sood[1,2], Charles Sievers[3,#], Yong Cheol Shin[1,#], Victoria Chen[1], Shunda Chen[3], Kirby K. H. Smithe[1], Sukti Chatterjee[4], Davide Donadio[3,*], Kenneth E. Goodson[5,6], Eric Pop[1,6,*]

[1]*Department of Electrical Engineering, Stanford University, Stanford CA 94305*
[2]*Present address: Stanford Institute for Materials and Energy Sciences, SLAC National Accelerator Laboratory, Menlo Park CA 94025*
[3]*Department of Chemistry, University of California, Davis CA 95616*
[4]*Applied Materials Inc., Santa Clara CA 95054*
[5]*Department of Mechanical Engineering, Stanford University, Stanford CA 94305*
[6]*Department of Materials Science and Engineering, Stanford University, Stanford CA 94305*
[#]equal contribution
[*]Corresponding authors: ddonadio@ucdavis.edu, epop@stanford.edu

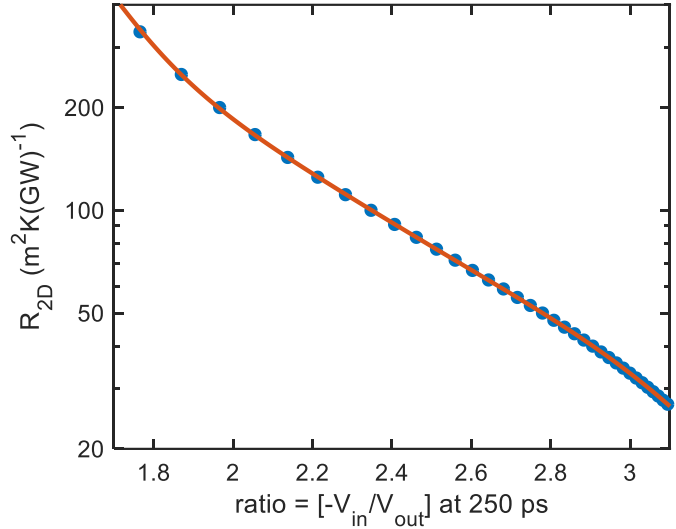

**Figure S1:** Transfer function relating the TDTR ratio signal (= $-V_{in}/V_{out}$) at +250 ps to the cross-plane thermal resistance of the 2D stack at the Al/SiO$_2$ interface, $R_{2D}$. Blue points are obtained from a solution to the 3D multilayer thermal model, while the red curve is a fit to a 6$^{th}$ order polynomial. The multilayer stack, from top to bottom, is: Al (80 nm)/{2D stack}/SiO$_2$ (90 nm)/Si.

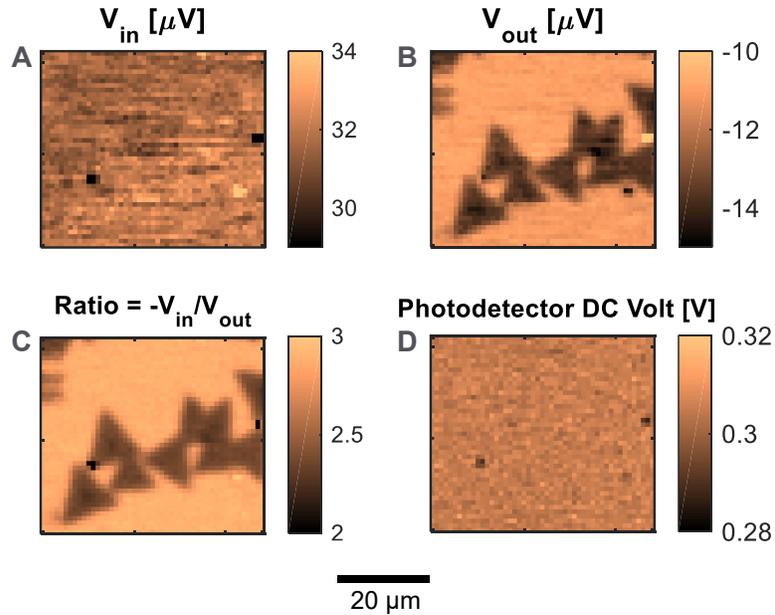

**Figure S2:** Maps of raw TDTR signals showing the (A) in-phase voltage $V_{in}$, (B) out-of-phase voltage $V_{out}$, (C) ratio = $-V_{in}/V_{out}$, and (D) DC probe reflectivity measured by the photodetector. Signals (A)-(C) are at a probe delay time of +250 ps. Because the sample is coated by an optically-opaque Al transducer layer, the DC probe reflectivity is uniform. Variations in the thermal resistance of the 2D interface between Al and SiO$_2$ appear largely as variations in the out-of-phase voltage (and thus the ratio). The ratio signal is converted to thermal resistance of the 2D interface using the transfer curve shown in **Fig. S1**.

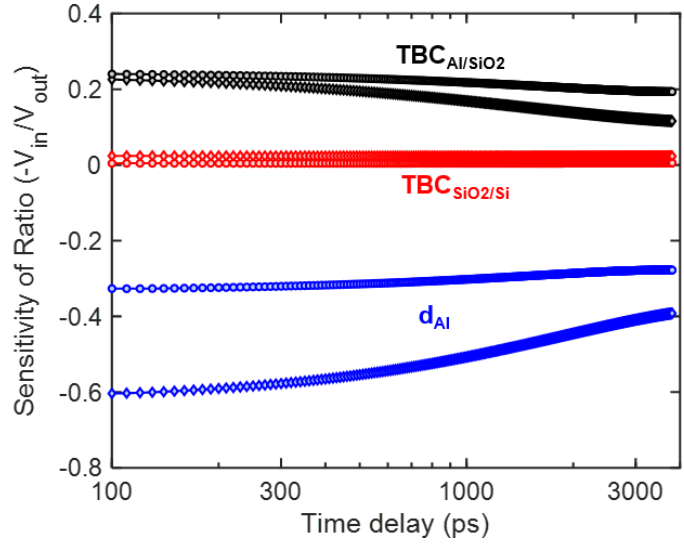

**Figure S3:** Sensitivity of the TDTR ratio signal to various parameters. TBC is the thermal boundary conductance = (thermal boundary resistance)$^{-1}$. The sensitivity coefficient for a parameter $\alpha$ is calculated as: $S_\alpha = \partial\log(Ratio)/\partial\log(\alpha)$, where $Ratio = -V_{in}/V_{out}$. Here, we examine the sensitivity to three parameters: (1) TBC at the Al/SiO$_2$ interface, which is equal to $(R_{2D})^{-1}$, and is the quantity we are interested in measuring (black markers), (2) TBC at the SiO$_2$/Si interface (red markers), and (3) Al transducer thickness (blue markers). Two sets of curves are plotted for extreme values of $R_{2D}$: diamonds for $R_{2D}$ = 50 m$^2$K(GW)$^{-1}$, circles for $R_{2D}$ = 200 m$^2$K(GW)$^{-1}$.

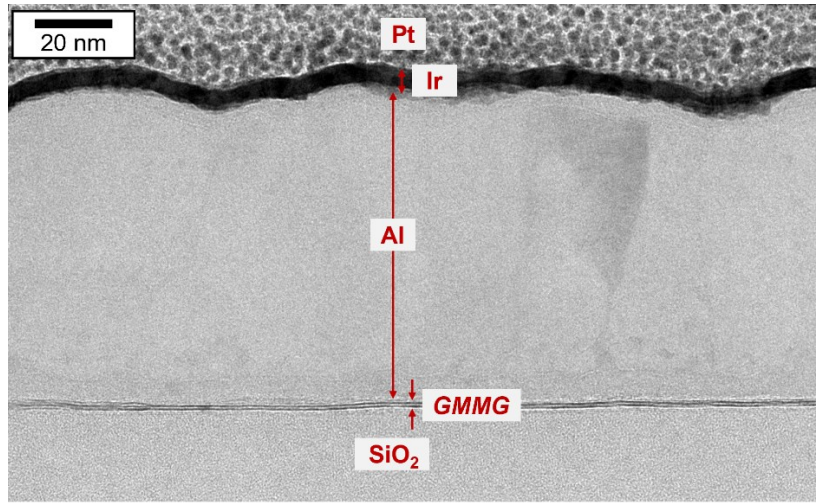

**Figure S4:** Cross sectional transmission electron micrograph (TEM) of a *GMMG* region in sample D2. The thickness of the Al transducer is 80 ± 2 nm (error bars based on the root-mean-square variation in thickness). The Ir and Pt layers are deposited during TEM sample preparation.

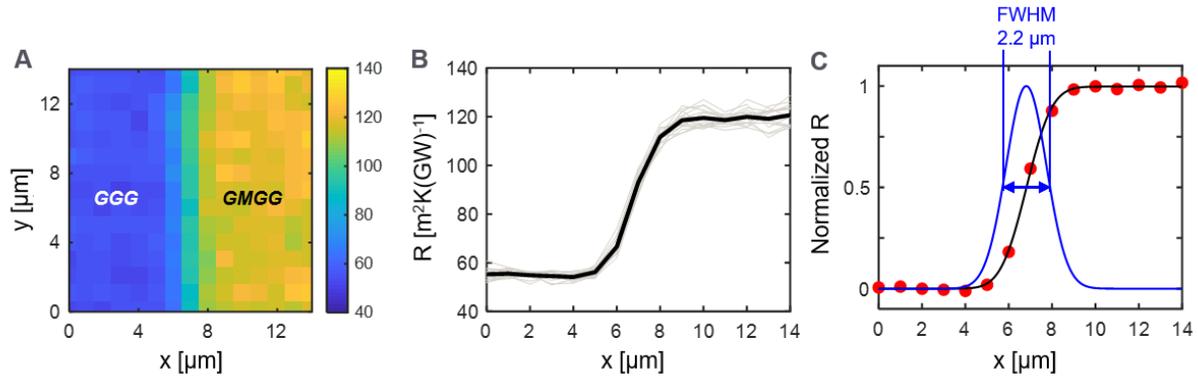

**Figure S5:** Spatial resolution of TDTR thermal resistance microscopy. (A) Thermal resistance map across a sharp junction between regions of the sample with *GGG* and *GMGG* stacking, i.e. the region on the right side has an additional $MoS_2$ monolayer inserted between the first two graphene layers. (B) Horizontal line cuts taken at 15 locations on the map (light grey) and the average of these line cuts (solid black). (C) Error function fit (black curve) of the average line cut data (red markers). Derivative of the error function curve gives a Gaussian (blue curve) with full-width half-maximum (FWHM) spatial resolution of 2.2 µm.

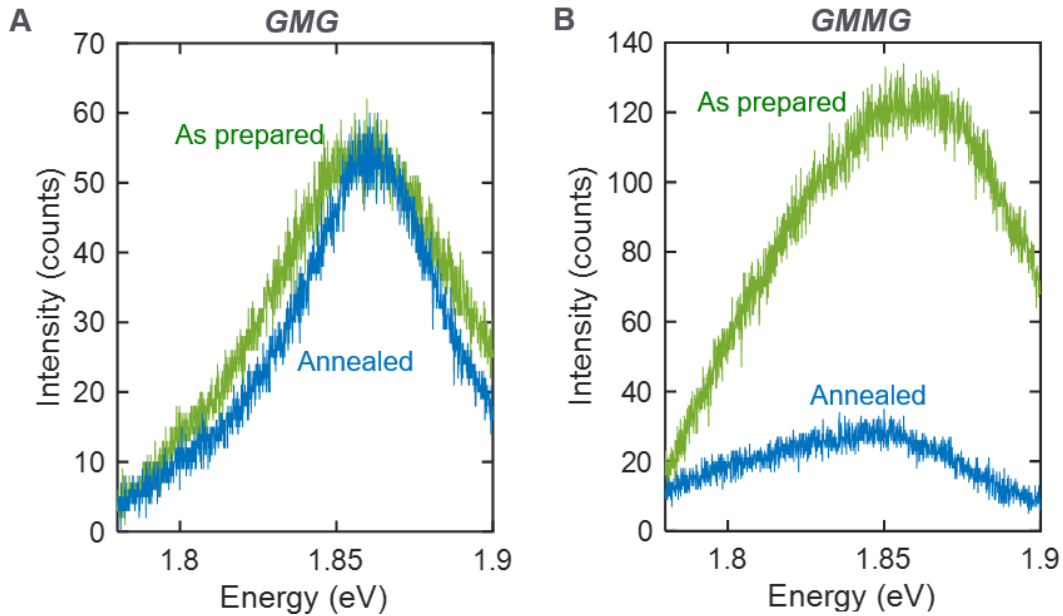

**Figure S6:** $MoS_2$ photoluminescence (PL) point spectra of (A) *GMG* and (B) *GMMG* regions of sample D2, showing the effect of annealing at 350°C for 3 hours under high vacuum (7 µTorr). No PL quenching is observed in *GMG* suggesting that annealing does not significantly modify the electronic coupling between graphene and $MoS_2$. In *GMMG*, the PL intensity is initially twice that in *GMG*, and is quenched upon annealing, due to the enhanced electronic coupling between the two $MoS_2$ monolayers.

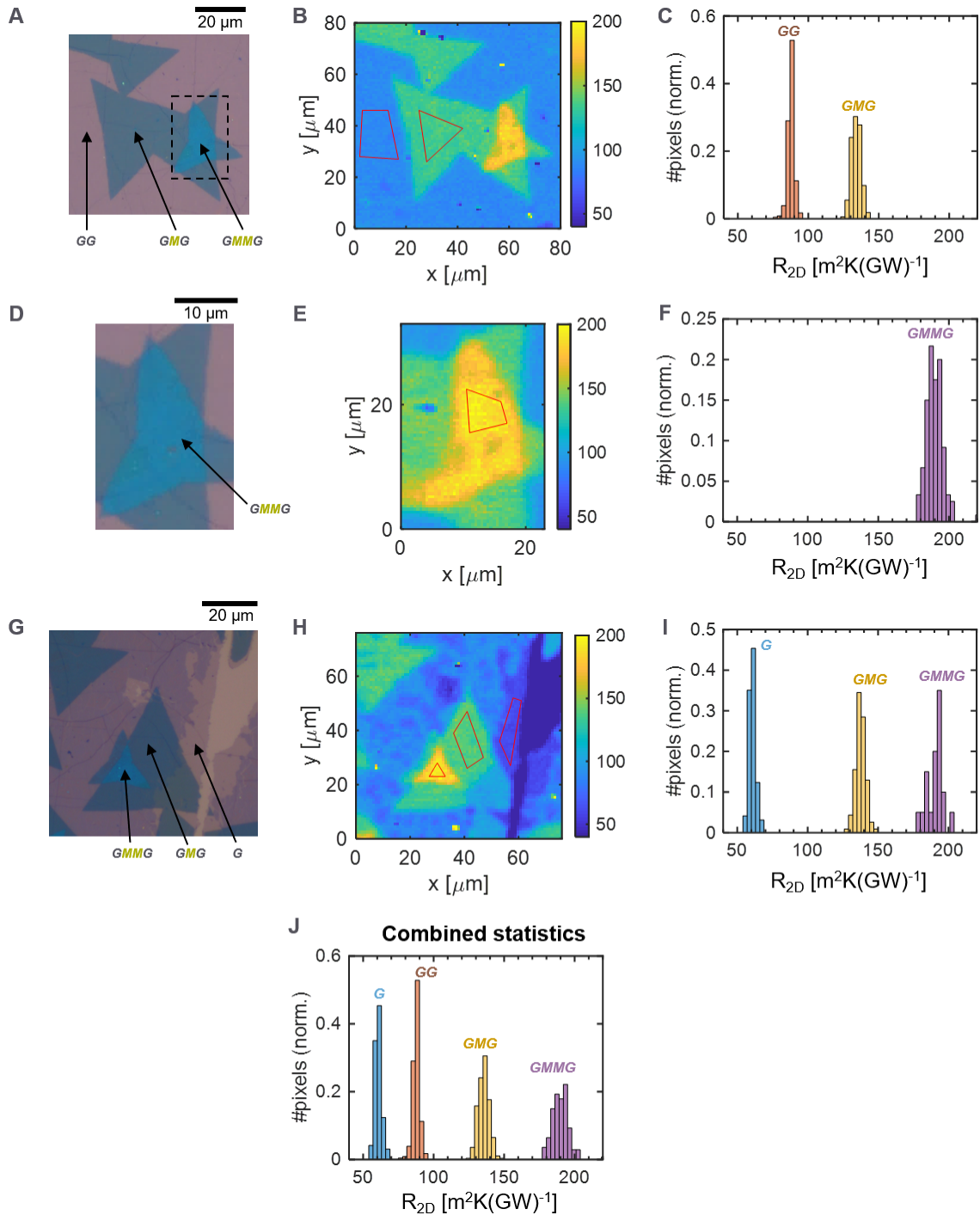

**Figure S7:** Comprehensive data on sample D1, showing optical micrographs, TDTR thermal resistance maps, and statistics. Regions of interest (ROIs) based on which the histograms are plotted are marked by the red polygons on the TDTR maps. (A)-(C) Region 1, providing data on *GG* and *GMG* regions. (D)-(F) Region 2, which zooms into the area enclosed within the dashed lines in (A), providing high-resolution data on *GMMG*. This TDTR map is measured with a step size of 500 nm. (G)-(I) Region 3, providing data on *G*, *GMG*, and *GMMG*. (J) Combined statistics taken from all 3 regions. In (C), (F), (I), and (J), each distribution is normalized by the number of pixels in the ROI.

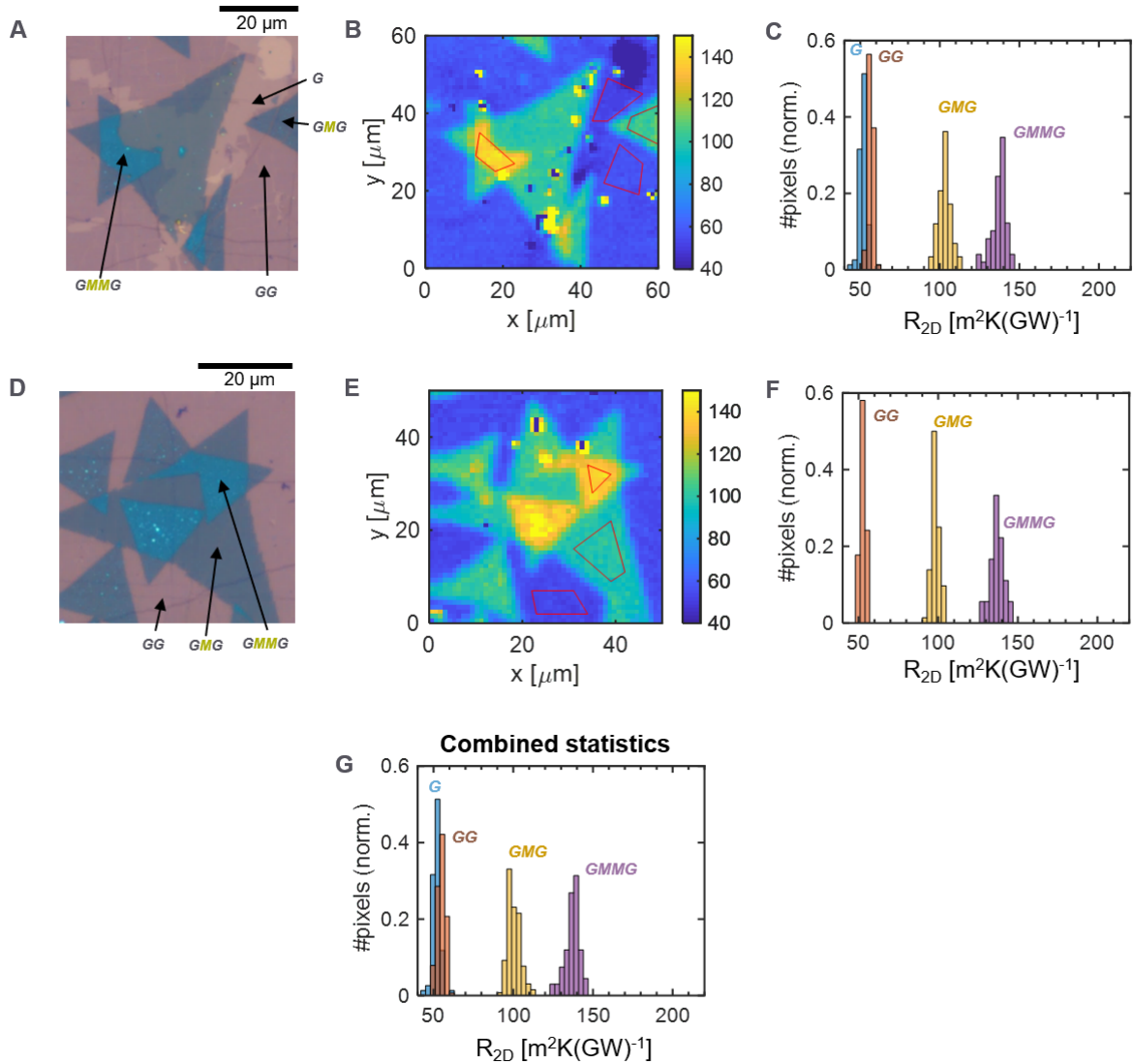

**Figure S8:** Comprehensive data on sample D2, showing optical micrographs, TDTR thermal resistance maps, and statistics. Regions of interest (ROIs) based on which the histograms are plotted are marked by the red polygons on the TDTR maps. (A)-(C) Region 1, providing data on *G*, *GG*, *GMG* and *GMMG* regions. (D)-(F) Region 2, providing data on *GG*, *GMG* and *GMMG* regions. (G) Combined statistics taken from both regions. In (C), (F), and (G), each distribution is normalized by the number of pixels in the ROI.

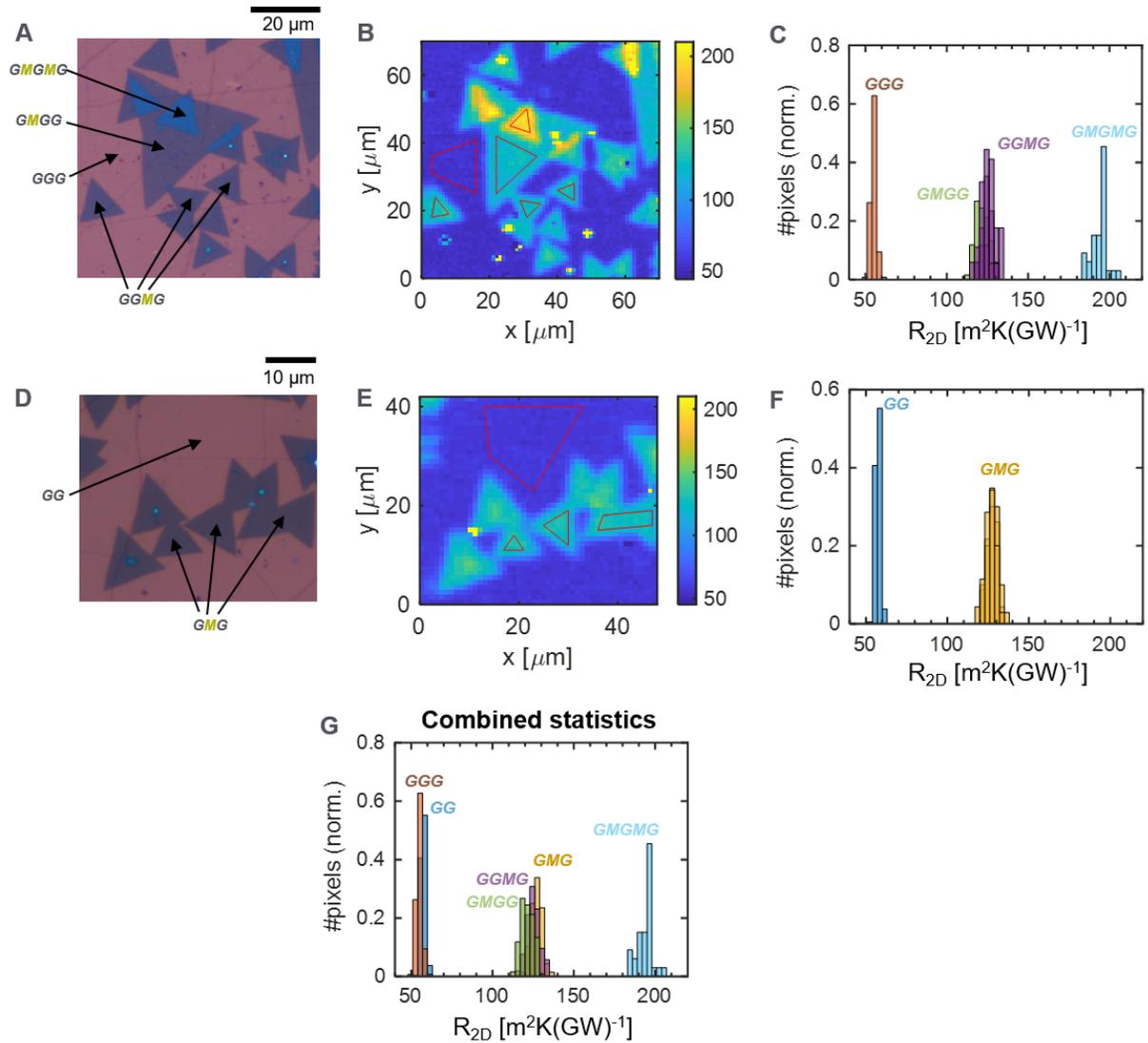

**Figure S9:** Comprehensive data on sample D3, showing optical micrographs, TDTR thermal resistance maps, and statistics. Regions of interest (ROIs) based on which the histograms are plotted are marked by the red polygons on the TDTR maps. (A)-(C) Region 1, providing data on *GGG*, *GGMG*, *GMGG*, and *GMGMG* regions. (D)-(F) Region 2, providing data on *GG* and *GMG* regions. (G) Combined statistics taken from both regions. In (C), (F), and (G), each distribution is normalized by the number of pixels in the ROI.

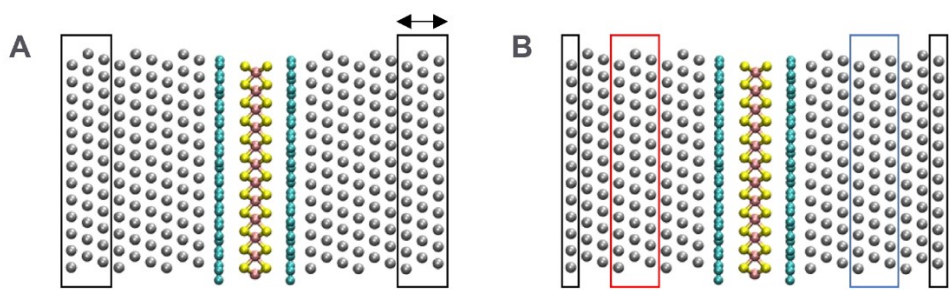

**Figure S10:** Simulation configurations. Grey atoms are aluminum, cyan atoms are carbon, pink atoms are molybdenum, yellow atoms are sulfur. Black boxes represent fixed atoms during calculations. (A) During QHA FEM the outer three layers of aluminum are fixed. The set of fixed atoms to the right of the device are displaced relative to the rest of the device to relax the cell (indicated by a double-headed arrow). (B) During NEMD, the outermost layer of aluminum on either side is fixed. The atoms in the red box represent the heat source which was set to 350 K and the blue atoms are the heat sink which are thermostatted to 250 K using a Langevin thermostat. The rest of the device was run in the NVE ensemble.

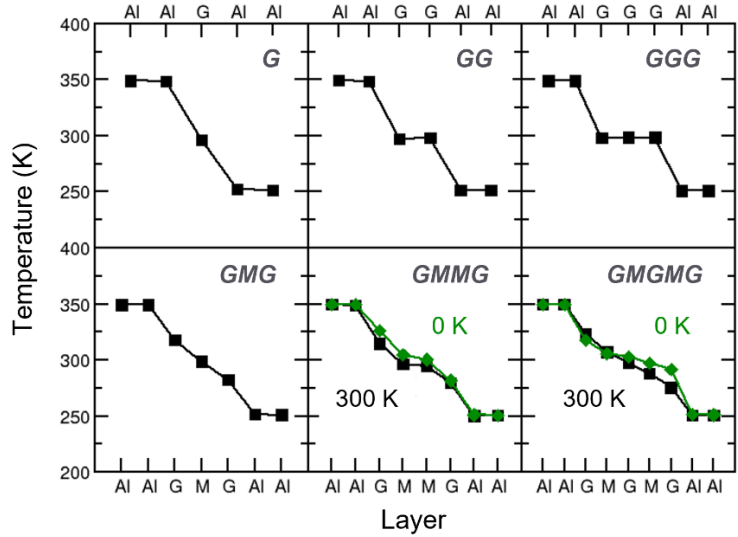

**Figure S11:** Temperature profiles averaged over the last 500 ps of each NEMD simulation (with 10 fs intervals). Each device shows large jumps in the temperature profile at the Al-*G* interfaces indicating a larger resistance at these interfaces. The *GG* and *GGG* systems exhibit relatively low *G-G* interfacial resistances compared to the Al-*G* interfaces, which is to be expected of homo *vs*. heterojunctions. The *GMMG* shows that the *M-M* homojunction exhibits lesser resistance than either the Al-*G* or *G-M* heterojunctions. The *GMMG* and *GMGMG* devices also have their 0 K optimized structures run at the same NEMD conditions for reference to see the effects of thermal expansion. We see that for *GMGMG*, as the device expands, the resistances of the *G-M* interfaces grow relative to the Al-*G* interfaces, however, for *GMMG* this is not as clearly the case. As the devices expand, the *G* and *M* layers become thermally decoupled from one another, which contributes to an increase in the overall thermal resistance.

**Table S1**: Error propagation analysis

| Stack | $\Delta_1$, error due to ±2 nm Al thick. (%) | $\Delta_2$, std. dev. in histograms (%) | $\Delta_3$, sample-to-sample var. (%) | Total error, $\Delta$ (%) |
|---|---|---|---|---|
| G | 7.0 | 4.4 | - | 8.3 |
| GG | 7.0 | 4.5 | 2 | 8.6 |
| GGG | 7.0 | 3.2 | - | 7.7 |
| GMG | 4.3 | 4.1 | 7 | 9.2 |
| GGMG | 4.3 | 2.9 | - | 5.2 |
| GMGG | 4.3 | 3.2 | - | 5.4 |
| GMMG | 3.7 | 3.2 | - | 4.9 |
| GMGMG | 2.7 | 2.5 | - | 3.7 |

Three sources of error are considered: (1) Caused by uncertainties in the thickness of the Al transducer, ±2 nm. (2) Spatial variations within a sample, determined from the standard deviation of the thermal resistance histograms. (3) Sample-to-sample variation, only relevant for *GG* and *GMG* where the same stack is found in more than one sample (D2 and D3). The total error is obtained by adding up the individual contributions in quadrature, $\Delta = (\Delta_1^2 + \Delta_2^2 + \Delta_3^2)^{1/2}$.